\documentclass[aps,prd,preprint,groupedaddress,longbibliography]{revtex4-1}

\usepackage{amsmath,amssymb,amsfonts}
\usepackage[T1]{fontenc}
\usepackage{graphicx}
\usepackage{adjustbox}
\usepackage{array}
\newcolumntype{L}{>{\centering\arraybackslash}p{3.2cm}}
\usepackage{hyperref}

\begin{document}

\title{Laboratory tests of the ordinary-mirror particle oscillations and the extended CKM matrix}


\author{Wanpeng Tan}
\email[]{wtan@nd.edu}
\affiliation{Department of Physics, Institute for Structure and Nuclear Astrophysics (ISNAP), and Joint Institute for Nuclear Astrophysics - Center for the Evolution of Elements (JINA-CEE), University of Notre Dame, Notre Dame, Indiana 46556, USA}

\date{\today}

\begin{abstract}
The CKM matrix and its unitarity is analyzed by disentangling experimental information obtained from three different particle systems of neutrons, mesons, and nuclei. New physics beyond the Standard Model is supported under the new analysis. In particular, the newly proposed mirror-matter model [Phys. Lett. B 797, 134921 (2019)] can provide the missing physics and naturally extend the CKM matrix. Laboratory experiments with current best technology for measuring neutron, meson, and nuclear decays under various scenarios are proposed. Such measurements can provide stringent tests of the new model and the extended CKM matrix.
\end{abstract}


\maketitle

\section{Introduction\label{intro}}

The Cabibbo–Kobayashi–Maskawa (CKM) matrix \cite{cabibbo1963,kobayashi1973} defines the strength of quark mixing in the standard model (SM). The CKM matrix for three families of quarks can be written as follows,
\begin{equation}
V_{CKM} = 
\begin{pmatrix}
V_{ud} & V_{us} & V_{ub} \\
V_{cd} & V_{cs} & V_{cb} \\
V_{td} & V_{ts} & V_{tb} \\
\end{pmatrix}
\end{equation}
which, under the unitarity condition, is fully defined by four independent parameters including a phase allowing for the only $CP$-violation effect \cite{bigi2009} confirmed in SM. More details can be found in a recent review \cite{porter2016}.

The unitarity of the CKM matrix and its matrix elements, in particular, $V_{ud}$, have been studied with various experimental efforts. However, inconsistent results have been reported from the three different types of decay measurements:
\begin{enumerate}
\item neutron decay or lifetime,
\item $K/\pi$ meson decays,
\item and nuclear transitions (superallowed $0^+ \rightarrow 0^+$ decays).
\end{enumerate}
The discrepancies may lie in the different properties of new physics manifested in three different particle systems of neutrons, kaons/pions, and nuclei, as discussed later. In order to disentangle such effects, different experiments using different particle systems for determining the matrix elements are separated in the following discussions. In particular, $V_{ud}$ derived from the superallowed $0^+ \rightarrow 0^+$ decays \cite{hardy2015} was often mixed with other measurements for evaluation of other matrix elements, which is avoided here.

Based on the newly proposed mirror-matter model \cite{tan2019}, we will try to reconcile all the major data sets for the three particle systems. In particular, we do not discredit any of the discussed data and assume no systematic or other experimental defects (unrelated to new physics) in the measurements. Then we demonstrate that all experimental data sets are consistent under the consideration of the new mirror-matter model. The apparent discrepancies are reasonably explained under the same framework.

\section{Analysis of the CKM Matrix\label{CKM}}

The matrix element $|V_{ub}| = 0.00394(36)$ \cite{particledatagroup2018} is very small and therefore it contributes little in studies of the unitarity.
The best direct constraint on $V_{us}$ is by measurements of the semileptonic $K_{l3}$ decays via
$f_+(0)|V_{us}| = 0.21654(41)$ \cite{moulson2017}
where the form factor at zero momentum transfer $f_+(0)$ is calculated to be $0.9696(15)_{stat}(11)_{sys}$ by the lattice QCD approach \cite{bazavov2018}. The best value for the matrix element $V_{us}$ is then,
\begin{equation} \label{eq_vus}
|V_{us}| = 0.22333(60).
\end{equation}
Hadronic $\tau$ decay experiments provide an independent measurement of $V_{us}$ which, however, uses $V_{ud}(0^+ \rightarrow 0^+)$ and has a larger uncertainty \cite{grouphflav2017}. Therefore, it is not considered here.

The ratio of the radiative inclusive rates for $K^{\pm}_{\mu 2}$ and $\pi^{\pm}_{\mu 2}$ decays sets
$f_{K^{\pm}}/f_{\pi^{\pm}} |V_{us}/V_{ud}| = 0.27599(37)$ \cite{moulson2017}
where the FLAG averaged lattice QCD calculations give the ratio of the isospin-broken decay constants $f_{K^{\pm}}/f_{\pi^{\pm}} = 1.1932(19)$ \cite{aoki2019}. The best value using the most recent updates is therefore,
\begin{equation} \label{eq_vusvud}
|V_{us}/V_{ud}| = 0.23130(48).
\end{equation}
The matrix element $V_{ud}$ can then be obtained from measurements of meson decays using Eqs. (\ref{eq_vus}-\ref{eq_vusvud}),
\begin{equation} \label{eq_meson}
|V_{ud}| = 0.9655(33).
\end{equation}

The PIBETA experiment by measuring the rare $\pi^+ \rightarrow \pi^0 e^+ \nu$ decay branching ratio offered a different meson $V_{ud}$ value of $|V_{ud}| = 0.9728(30)$ \cite{pocanic2004}.
Later we will show that this measurement is consistent with either the proposed new physics or assumption of no new physics due to its large uncertainty.

For neutron $\beta$ decays, the matrix element $V_{ud}$ can be written as,
\begin{eqnarray} \label{eq_n}
|V_{ud}|^2 &=& \frac{2\pi^3}{G^2_F m^5_e f_n \tau_n (1+3\lambda^2)(1+\delta'_R)(1+\Delta^V_R)} \nonumber \\
&=& \frac{5024.46(30) \text{ sec}}{\tau_n (1+3\lambda^2)(1+\Delta^V_R)}
\end{eqnarray}
where the Fermi constant $G_F=1.1663787(6)\times 10^{-5}$ GeV$^{-2}$, $m_e$ is the electron mass, the neutron-specific radiative correction $\delta'_R = 0.014902(2)$ \cite{towner2010}, the phase space factor $f_n$ is 1.6887(1) \cite{towner2010,czarnecki2018}, and natural units ($\hbar=c=1$) are used for simplicity. The 1\% difference in neutron $\beta$-decay lifetime $\tau_n$ between measurements from ``beam'' and ``bottle'' experiments leads to the discrepant $V_{ud}$ values according to Eq. (\ref{eq_n}). The neutron lifetime anomaly becomes more severe by more than $4\sigma$ from recent high-precision measurements \cite{yue2013,pattie2018}.

More recent measurements on the ratio of the axial-to-vector couplings $\lambda = g_A/g_V$ especially after 2002 have provided more reliable values \cite{czarnecki2018} and its current best value of $\lambda = -1.27641(56)$ comes from the PERKEO III measurement \cite{markisch2018}. One of the largest uncertainties other than the neutron lifetime in Eq. (\ref{eq_n}) is from the transition-independent radiative correction and its newly updated value is $\Delta^V_R=0.02467(22)$ \cite{seng2018}. Using the neutron $\beta$-decay lifetime of $\tau_n = 888.0\pm 2.0$ s from the averaged ``beam'' values \cite{byrne1996,yue2013}, we can obtain the matrix element,
\begin{equation} \label{eq_beam}
|V_{ud}| = 0.9684(12).
\end{equation}

The ``bottle'' lifetime measurements using ultra-cold neutrons (UCN) are not consistent within themselves possibly due to the differences of the trap geometry \cite{tan2019}. For example, the two most recent measurements \cite{pattie2018,serebrov2018} deviate from each other by $3.2\sigma$. Using the average ``bottle'' lifetime of $\tau_n(\text{bottle}) = 879.4 \pm 0.6$ s, we can obtain $|V_{ud}(\text{bottle})| = 0.97317(50)$ from Eq. (\ref{eq_n}).

The matrix element determined from the superallowed $0^+ \rightarrow 0^+$ decays \cite{hardy2015} using the updated $\Delta^V_R$ value \cite{seng2018} is $|V_{ud}(0^+ \rightarrow 0^+)| = 0.97370(14)$ \cite{particledatagroup2018}, which disagrees with the ``beam'' value and also exhibits a tension with $V_{ud}$ in Eq. (\ref{eq_meson}) from the meson decay measurements. Using $V_{us}$ from Eq. (\ref{eq_vus}), unitarity of the CKM matrix is violated for any of the above-discussed $V_{ud}$ values (except for the PIBETA data) as shown in Table \ref{tab_1}. In particular, the deviation is $5.3\sigma$ for the most trusted $V_{ud}(0^+ \rightarrow 0^+)$, indicating that new physics is needed.

\begin{table}
\caption{\label{tab_1} Deviation in $\sigma$-level from unitarity of the CKM matrix for the first row of $|V_u|^2 = |V_{ud}|^2 + |V_{us}|^2 + |V_{ub}|^2$ is shown based on $|V_{us}|=0.22333(60)$ from Eq. (\ref{eq_vus}) , $|V_{ub}| = 0.00394(36)$ \cite{particledatagroup2018}, and different $V_{ud}$ values from different types of decay measurements.}
\begin{ruledtabular}
\begin{tabular}{c c c c}
Measurement Type & $|V_{ud}|$ & $|V_u|^2$ & $\sigma$ \\
\hline
meson ($K_{l3}$) & 0.9655(33) & 0.9822(63) & 2.8 \\
meson (PIBETA) & 0.9728(30) & 0.9962(66) & 0.6 \\
n-decay ``beam'' & 0.9684(12) & 0.9878(22) & 5.4 \\
n-decay ``bottle'' & 0.97317(50) & 0.9969(10) & 3.0 \\
nuclear $0^+ \rightarrow 0^+$ & 0.97370(14) & 0.99798(38) & 5.3 \\
\end{tabular}
\end{ruledtabular}
\end{table}

The discrepancy of $V_{ud}$ values between meson ($K_{l3}$) and nuclear decay measurements has not drawn as much attention as the neutron lifetime anomaly, partly due to still larger uncertainties in meson decay studies. Another reason is that $V_{ud}(0^+ \rightarrow 0^+)$ is often treated as the gold standard for obtaining other matrix elements. $V_{ud}$ derived from the superallowed $0^+ \rightarrow 0^+$ decays is so trusted that exclusion of any exotic decay channels of neutrons was proposed \cite{dubbers2019}. As for the neutron lifetime anomaly, the ``bottle'' method has become more favored for obtaining the neutron $\beta$-decay lifetime owing to its agreement with $V_{ud}(0^+ \rightarrow 0^+)$ or its apparent consistency with the axial-to-vector coupling ratio $\lambda$ derived from recent $\beta$-asymmetry measurements as studied in Ref. \cite{czarnecki2018}. However, the tension with $V_{ud}$ inferred from measurements of meson decays ($K_{l3}$ and $K_{\mu 2}$/$\pi_{\mu 2}$) may reverse all these arguments, which will be discussed in the next section.

Assuming that $V_{ud}(0^+ \rightarrow 0^+)$ is the standard with no new physics, one can find in Table \ref{tab_1} that the $V_{ud}$ values derived from the $K_{l3}$ data and the ``beam'' approach deviate from $V_{ud}(0^+ \rightarrow 0^+)$ by $2.5\sigma$ and $4.5\sigma$, respectively. Furthermore, the unitarity requirement can make the tension go beyond the $5\sigma$ level as shown in Table \ref{tab_1}. As a result, one has to discredit two independent data sets of $K_{l3}$ decays and ``beam'' lifetime assuming no new physics.

On the contrary, with the new physics from the particle-mirror particle oscillations \cite{tan2019} that will be presented in the next section, all these data sets can be reconciled and the apparent discrepancies can be explained. In particular, the smaller ``bottle'' lifetime values are due to the loss from $n-n'$ oscillations \cite{tan2019} and the larger $V_{ud}$ from the superallowed beta decays is resulted from unaccounted radiative corrections of the new physics as discussed later. In this case, $V_{ud}$ is best determined by the ``beam'' approach, which is consistent with both meson data sets ($<1\sigma$ for $K_{l3}$ and $\sim 1.3\sigma$ for PIBETA).

\section{Mirror-Matter Model and the Extended CKM Matrix\label{model}}

Various theoretic efforts \cite{berezhiani2006,fornal2018,berezhiani2019a,tan2019} have been devoted for solving the issues, in particular, the neutron lifetime anomaly. The idea of neutron dark decay in nuclei \cite{pfutzner2018} based on the dark decay model of Fornal and Grinstein \cite{fornal2018} pointed to clues of new physics from nuclear systems. A 4th quark in the mixing with the three known generations of quarks was recently suggested to solve the discrepancies in the CKM unitarity \cite{belfatto2020}. Most of the previous works focus on correcting the ``beam'' lifetime to agree with the ``bottle'' lifetime. Some of the more interesting models introduce $n-n'$ oscillations involving the mirror-matter theory \cite{berezhiani2006,berezhiani2019a,tan2019}.

There are several models on the mirror matter theory that have been proposed. Typically very weak interactions besides gravity between particles of ordinary and mirror sectors are introduced. A photon-mirror photon kinetic mixing mechanism was suggested to couple the two sectors \cite{holdom1986,foot2004}. The possibility of transition magnetic moments between the ordinary and mirror neutrons was studied as well \cite{berezhiani2018b}. Alternatively, a six-quark coupling was induced for the mixing of ordinary and mirror neutrons and explanation of the neutron lifetime anomaly \cite{berezhiani2006}.

Spontaneous symmetry breaking of the mirror symmetry was also used in various degrees. It was first used for an unsuccessful attempt of explanation of neutrino oscillations \cite{berezhiani1995} by introducing a mirror symmetry breaking scale of a factor of 30. The idea was also applied to mirror matter theory in general for extensive exploration of issues in particle physics and cosmology \cite{cui2012}. To reconcile the neutron lifetime discrepancies, an $n-n'$ oscillation model was proposed using a six-quark coupling and a small $n-n'$ mass splitting of $10^{-7}$ eV \cite{berezhiani2019a} where, like many other studies, the ``bottle'' lifetime is favored again.

Different models aim at different ways to solve the above discrepancies. In the following, we will introduce briefly the newly proposed mirror-matter model \cite{tan2019} and discuss various laboratory tests that can be carried out with current technology and distinguish this model from other proposed solutions.

In this rather exact new mirror matter model \cite{tan2019}, no cross-sector interaction is introduced, unlike other models. It can consistently and quantitatively explain various observations in the Universe including dark energy \cite{tan2019e}, the neutron lifetime puzzle and dark-to-baryon matter ratio \cite{tan2019}, origin of baryon asymmetry \cite{tan2019c}, evolution and nucleosynthesis in stars \cite{tan2019a}, and ultrahigh energy cosmic rays \cite{tan2019b}. Extension of the model into a hierarchy of supersymmetric mirror models has been developed to explain the dynamic evolution of the Universe and underlying physics \cite{tan2019e,tan2020,tan2020a} and to understand the nature of black holes \cite{tan2020b}.

The critical assumption of this model is that the mirror symmetry is spontaneously broken by the uneven Higgs vacuum in the two sectors, i.e., $<\phi> \neq <\phi'>$, although very slightly (on a relative breaking scale of $10^{-15} \text{--} 10^{-14}$) \cite{tan2019}. The breaking of the mirror symmetry is supported by a new theorem stating that no global symmetries in quantum gravity are possible \cite{harlow2019}. When fermion particles obtain their mass from the Yukawa coupling, it automatically leads to the mirror mixing for neutral particles, i.e., the basis of mass eigenstates is not the same as that of mirror eigenstates, similar to the generation mixing of quarks and neutrinos. Further details of the model can be found in Ref. \cite{tan2019}.

The immediate result of this model is the probability of ordinary-mirror neutral particle oscillations in vacuum \cite{tan2019},
\begin{equation}\label{eq_prob}
P(t) = \sin^2(2\theta) \sin^2(\frac{1}{2}\Delta t)
\end{equation}
where $\theta$ is the mixing angle, $\sin^2(2\theta)$ denotes the mixing strength of about $10^{-4}$ for $K^0-K^{0'}$ and $2\times 10^{-5}$ for $n-n'$ with a possible range of $8\times10^{-6}$ - $4\times10^{-5}$, $t$ is the propagation time, and $\Delta$ is the small mass difference of the two mass eigenstates (on the order of $10^{-6}$ eV for both $K^0-K^{0'}$ and $n-n'$) \cite{tan2019}.

Under the new model, the symmetry breaking may occur in the same way \cite{tan2019c} for the two discrete family ($Z_3$) and mirror ($Z_2$) symmetries resulting in one extended quark mixing matrix as follows,
\begin{equation} \label{eq_qmix}
V_{qmix} = 
\begin{pmatrix}
V_{ud} & V_{us} & V_{ub} & V_{uu'} \\
V_{cd} & V_{cs} & V_{cb} & V_{cc'} \\
V_{td} & V_{ts} & V_{tb} & V_{tt'} \\
V_{dd'} & V_{ss'} & V_{bb'} & V' \\
\end{pmatrix}
\end{equation}
where the quark-mirror quark mixing elements $V_{qq'}$ could, as a naive estimate, be very similar to each other. For simplicity, $V'$ represents the $3\times3$ CKM matrix within the mirror sector and other cross-sector elements are assumed to vanish at least in the first order and hence suppressed. The unitarity condition for the first row of the matrix can then be written as $|V_{ud}|^2+ |V_{us}|^2+ |V_{ub}|^2+ |V_{uu'}|^2=1$. As shown in Table \ref{tab_2}, this results in $|V_{uu'}| \simeq 0.11$ using the ``beam'' value of $V_{ud}$ from Eq. (\ref{eq_beam}) and $V_{us}$ from Eq. (\ref{eq_vus}).

\begin{table}
\caption{\label{tab_2} Elements of the extended mixing matrix in Eq. (\ref{eq_qmix}) adopted or predicted under the new mirror-matter model assuming the unitarity condition. In particular, $V_{ud}$ is taken from the ``beam'' lifetime approach while $V_{us}$ is from Eq. (\ref{eq_vus}) on $K_{l3}$ decays and $V_{ub}$ is from Ref. \cite{particledatagroup2018}. The cross-sector elements are predicted using the $n-n'$ and $K^0-K^{0'}$ mixing strengths from Refs. \cite{tan2019,tan2019c}.}
\begin{ruledtabular}
\begin{tabular}{c c c c c c}
$|V_{ud}|$ & $|V_{us}|$ & $|V_{ub}|$ & $|V_{uu'}|$ & $|V_{dd'}|$ & $|V_{ss'}|$\\
\hline
0.9684(12) & 0.22333(60) & 0.00394(36) & 0.11 & 0.071 & 0.035 \\
\end{tabular}
\end{ruledtabular}
\end{table}

Note that quarks and mirror quarks obey different gauge symmetries within their own sectors and are thus charged differently. The result is that the $qq'$ mixing does not manifest at the single-quark level because of charge conservation. However, for neutral mesons and neutrons, this mixing effect does show up as oscillations like $K^0-K^{0'}$ \cite{tan2019c} and $n-n'$ \cite{tan2019}. Therefore, such neutral hadron oscillations (and likely neutrino oscillations) could be understood as a type of topological oscillations due to the broken mirror symmetry and mass splitting in contrast to other known nonperturbative transitions overcoming energy barriers such as instantons \cite{thooft1976,thooft1976a}, sphalerons \cite{klinkhamer1984}, and quarkitons \cite{tan2019c}.

The mixing strength of neutral particles that are made of quarks can then be obtained from the mirror-mixing matrix elements for all constituent quarks,
\begin{equation}\label{eq_mixv}
\sin^2(2\theta) \simeq \prod_{i} |2V_{q_iq'_i}|^2
\end{equation}
where the mixing angle $\theta$ is assumed to be small. Using Eq. (\ref{eq_mixv}), more mirror-mixing elements can be estimated with the known mixing strength. For example, the $n-n'$ mixing strength $\sin^2(2\theta_{nn'}) = |2V_{uu'}|^2|2V_{dd'}|^4 \simeq 2\times 10^{-5}$ leads to $|V_{dd'}| \simeq 0.071$. The study of $K^0_L-K^{0'}_L$ oscillations in the early universe for the origin of baryon asymmetry \cite{tan2019c} supports the mixing strength $\sin^2(2\theta_{KK'}) = |2V_{dd'}|^2|2V_{ss'}|^2 \simeq 10^{-4}$ resulting in $|V_{ss'}| \simeq 0.035$ as shown in Table \ref{tab_2}. The similarity in the amplitude of the mirror-mixing elements indicates that other relatively long-lived neutral hadrons (e.g.,  $\Lambda^0$, $\Xi^0$, $D^0$ and $B^0$) with lifetimes comparable to the oscillation time scale of nanoseconds could also exhibit significant oscillation effects. On the contrary, for the lightest mesons, $\pi^0-\pi^{0'}$ oscillations are negligible with a fraction of $< 10^{-18}$ due to the extremely short lifetime \cite{tan2019}.

$V_{ud}$ values determined from meson decays and the ``beam'' lifetime are consistent and support the new mirror-matter model. All the data sets discussed in the previous section can then be reconciled under the new model. The apparent consistency between the ``bottle'' lifetime and the superallowed $0^+ \rightarrow 0^+$ decays is probably accidental. The anomalous ``bottle'' lifetime is explained with the loss of neutrons via $n-n'$ oscillations when bouncing inside a trap \cite{tan2019}. The mixing elements $V_{uu'}$ and $V_{dd'}$ may introduce additional unaccounted radiative corrections (e.g., virtual $n-n'$ oscillations) to the superallowed $0^+ \rightarrow 0^+$ decays, which will lower calculated $V_{ud}(0^+ \rightarrow 0^+)$ accordingly. Due to energy conservation, the virtual oscillations can not become external in typical nuclei including the ones undergoing superallowed beta decays. On the other hand, such virtual oscillation processes will emerge as hidden decay branches in neutral meson and neutron decays that provide a cleaner way of determining matrix elements compared to nuclear decays. As will be presented in the next section, the virtual $n-n'$ oscillations could also manifest as invisible $\beta' p'$ decays and unexpectedly strong $\beta p$ decays in some so-called one-neutron halo nuclei. Early results seem to indicate the existence of such virtual $n-n'$ oscillations in nuclei.

\section{Laboratory Tests\label{test}}

As discussed above, $V_{ud}$ and $V_{us}$ determined from meson decays play a critical role. More accurate branching ratio measurements on the meson decays of $K_{l3}$, $K^{\pm}_{\mu 2}$ and $\pi^{\pm}_{\mu 2}$ are in need and meanwhile the corresponding hadronic constants $f_+(0)$ and $f_{K^{\pm}}/f_{\pi^{\pm}}$ need improvements in the lattice QCD calculations as their uncertainties are comparable with the experimental counterparts. This will provide an independent value of $V_{ud}$ in comparison with the $V_{ud}$ value from the neutron lifetime measurement of the ``beam'' approach. Similar advancement for the $\pi$ beta decay measurements should also be pursued. The agreement with much better uncertainties will provide very strong support for the new model. In addition, better $V_{ud}$ and $V_{us}$ values will define a better value of $V_{uu'}$ and further reveal the mechanism of the mirror-particle oscillations by fulfilling self-consistent checks on the single-quark mixing strengths inferred from $n-n'$ and $K^0-K^{0'}$ oscillations.

Under the new $n-n'$ oscillation model, the deviation of the neutron lifetime measured in the ``bottle'' approach from that in the ``beam'' method is due to the neutron loss by an averaged fraction of $\sin^2(2\theta)/2$ in each collision with the walls (or the confining magnetic fields for that matter) even if the wall surface itself is perfect \cite{tan2019}. Such an $n-n'$ oscillation mechanism results in a dependence of the measured lifetime on the geometry of a UCN trap. In particular, magnetic traps are better for such tests as imperfect wall surface conditions can be avoided. Experiments using traps with significantly different mean free flight times will provide one of the most stringent tests on the new model and distinguish it from other models. In particular, magnetic traps of a narrow cylindrical shape \cite{dzhosyuk2005,leung2016} could be re-run with better precision (e.g., close to that of the UCN$\tau$ measurement \cite{pattie2018}) providing an immediate test of the model. Upon confirmation of the model, these measurements will also nail down the mixing strength of $n-n'$ oscillations and therefore provide a better estimate of the matrix elements $V_{uu'}$ and $V_{dd'}$.

The other parameter of the new model is the mass splitting $\Delta$ between ordinary and mirror particles. Similar mass splitting values of the order of $10^{-6}$ eV for $\Delta_{nn'}$ and $\Delta_{K_L K_S}$ that accounts for the $CP$ violation in SM indicate that all these phenomena may stem from the same mechanism of spontaneous mirror symmetry breaking \cite{tan2019c} supporting the extended mixing matrix of Eq. (\ref{eq_qmix}). Under the consistent picture of the origin of both dark matter and baryon asymmetry \cite{tan2019c}, we can obtain $\Delta_{nn'} = 3\times 10^{-6}$ eV with the $n-n'$ mixing strength of $\sin^2(2\theta) = 2\times 10^{-5}$. If we assume that the mirror mass splitting scale is the same as that in $CP$ violation with $\Delta_{K_L K_S} = 3.484(6)\times 10^{-6}$ eV, we can derive a very precise $n-n'$ mass splitting value of $\Delta_{nn'}=6.578(11)\times 10^{-6}$ eV and the corresponding mixing strength of $\sin^2(2\theta) = 1.3\times 10^{-5}$. Further calculations based on this assumption have just been done for the study of invisible decays of long-live neutral hadrons \cite{tan2020d}.

Laboratory tests of the mass splitting parameter can be done with a setup similar to the ``beam'' approach \cite{nico2005} in neutron lifetime measurements. Neutrons in a magnetic field can be affected by an effective potential of $\mu B$ where $\mu = 6\times 10^{-8}$ eV/T is the absolute neutron magnetic moment \cite{tan2019}. The medium effect of a constant magnetic field can change the $n-n'$ oscillation probability to \cite{tan2019a}
\begin{equation}
P(t) = \frac{\sin^2(2\theta)}{C^2} \sin^2(\frac{1}{2}C\Delta_{nn'} t)
\end{equation}
where the medium factor $C^2 = (\cos(2\theta) - \mu B/\Delta)^2 + \sin^2(2\theta)$. Treatment for a varying field can be found in Ref. \cite{berezhiani2019a} as studied in detail for the matter effect of neutrino oscillations \cite{giunti2007a}.
Such an effect is negligible for typical magnetic fields of $B \lesssim 5$ T since $\mu B \ll \Delta_{nn'}$. Similar to the matter effect for neutrino oscillations and $n-n'$ oscillations in stars \cite{tan2019a}, however, the $n-n'$ oscillations can become resonant in very strong magnetic fields when $\mu B \sim \Delta_{nn'}$. For $\Delta_{nn'} = 3\times 10^{-6}$ eV, the resonant condition is $B=50$ T resulting in maximal mixing. Direct-current high fields up to 45.5 T have recently been demonstrated in a very compact magnet setup with new conductor material and a novel design \cite{hahn2019}. In a ``beam'' approach setup \cite{nico2005}, we could observe a significant neutron loss rate due to resonant $n-n'$ oscillations when the magnetic field is slowly ramped up to about 50 T using the new technology. For an unpolarized neutron beam, a simple estimate of the neuron loss gives 25\% when the resonant $n-n'$ mixing occurs. Such a large effect could simplify the ``beam'' setup significantly, i.e., not needing the detection of protons. Alternatively, pulsed magnets could be used to cover higher fields (e.g., 40-100 T) \cite{nguyen2016} in small steps (e.g., 0.1 T) for search of the resonant parameters should the $n-n'$ mixing strength is lower or $\Delta_{nn'}$ is larger. If it does confirm the new model, this laboratory test will help determine the mass splitting parameter more accurately. The sign of $\Delta_{nn'}$ is still unclear although application of this model to star evolution \cite{tan2019a} indicates that it is probably positive. Future experiments using polarized neutron beams under such super-strong magnetic fields may help determine the sign.

The resonant condition of the matter effect for $n-n'$ oscillations can be easily met at densities of $10^2-10^3$ g/cm$^3$ in stars \cite{tan2019a} while it is not feasible for a laboratory test on Earth. Nevertheless, we can conduct tests of cold/thermal neutrons traveling in a large detector made of dense and nearly absorption-free material. For example, cold neutron scattering occurs inside a liquid $^4$He detector at a temperature of 4 K with the following properties: neutron velocity $v = 2.5 \times 10^4$ cm/s; liquid $^4$He density $\rho = 0.125$ g/cm$^3$; scattering cross section $\sigma = 1.34$ b; and no absorption. From these parameters we can calculate a large ratio of the $n-n'$ oscillation rate to the ordinary neutron $\beta$-decay rate: $\lambda_{nn'} / \lambda_{\beta} \simeq 6$ under the new model \cite{tan2019}. The neutron mean free path in liquid $^4$He is $l = 40$ cm. To keep the collisions inside the detector within the oscillation time scale of $1/\lambda_{nn'} \sim 160$ s, however, the detector size has to be as large as $l/\sqrt{\sin^2(2\theta)/2} \sim 10^4$ cm. Fortunately, a smart design using magnetic fields to confine UCNs in a small volume of liquid $^4$He was realized for neutron lifetime measurements \cite{huffman2000,huffer2017}. An anomaly of $707\pm20$ s in the lifetime measurements was reported in an unpublished thesis work although the suspicious unexpected $^3$He contamination was blamed \cite{huffer2017}. Such an anomaly can be explained using the new model as follows: the $^4$He scattering is negligible at very low superfluid temperatures such that $^4$He serves as the detection medium only; the magnetic reflection rate could be about 30 s$^{-1}$ assuming the mean UCN velocity of $3$ m/s and the mean free path of 10 cm in the Ioffe trap; hence $n-n'$ oscillations result in an apparent lifetime of $\sim 700$ s. Further investigation using such a device would be a good test for the new model.

Another example is a heavy-water (D$_2$O) detector with neutrons at room temperature. We have the following parameters: scattering cross sections of $\sigma($D$) = 7.64$ b and $\sigma($O$) = 4.232$ b; absorption cross sections of $\sigma_{abs}($D$) = 5.19\times 10^{-4}$ b and $\sigma_{abs}($O$) = 1.9\times 10^{-4}$ b; density of $\rho($D$_2$O$) = 1.11$ g/cm$^3$; and neutron velocity of $v=2.2\times 10^5$ cm/s. The corresponding ratio of the $n-n'$ oscillation rate to the absorption rate, $\lambda_{nn'} / \lambda_{abs} \simeq 0.16$, is smaller compared to the case of liquid $^4$He. However, a much smaller mean free path of $l =1.5$ cm results in a much smaller detector size of about 5 meters in radius to contain neutrons within the oscillation time scale of one second.

Experiments measuring the branching fractions of $K^0_L$ and $K^0_S$ invisible decays can provide a different test (i.e., on $V_{ss'}$) of the new mirror-matter model \cite{tan2019}. Unfortunately, such branching fractions have not been constrained experimentally \cite{gninenko2015}. Under the new model assuming that the single quark mixing element $V_{qq'} \sim 0.1$, we can obtain an estimate of the branching fractions of $K^0$ invisible decays of about $10^{-6}$ for $K^0_S$ and $10^{-4}$ for $K^0_L$, which provides an explanation of the origin of baryon asymmetry in the early universe \cite{tan2019c} and is reachable with current experimental capabilities. Measurements of the branching fractions at current kaon production facilities can determine the mixing element $V_{ss'}$ more accurately and constrain the conditions of baryogenesis in the early universe. With future detector technology and accelerators, matrix elements of $V_{cc'}$ and $V_{bb'}$ could also be tested with similar measurements on the branching fractions of about $10^{-9}-10^{-10}$ for $D^0$ and $B^0$ invisible decays \cite{tan2019}. We may also have better chances of observing $\Lambda^0$ and $\Xi^0$ invisible decays at a branching fraction of $\sim 10^{-8}$ due to their long lifetimes. Better estimates of the invisible decay branching fractions of these neutral hadrons can be made assuming equivalence of the $CP$-violation and mirror symmetry breaking scales \cite{tan2020d}.

The $n-n'$ oscillation effects could also be studied in the quasi-free medium of a halo nucleus. The best example is $^{11}$Be, a one-neuron halo nucleus. It has a 13.76 second $\beta$-decay half-life with a strong $\beta$-delayed particle decay ($\beta\alpha$) branch  of 3.3\% \cite{refsgaard2019}. A rare decay branch of $^{11}$Be $\rightarrow$ $^{10}$Be was measured using the accelerator mass spectrometry (AMS) technique with an unexpectedly high branching ratio of $8.3\pm0.9 \times 10^{-6}$ \cite{riisager2014}. Considering the neutron separation energy $S_n = 501.64$ keV for $^{11}$Be, the neutron emission channel is not open while the $\beta$-delayed proton emission, i.e., the $\beta p$ decay, is energetically possible and the only possible process without new physics for $^{11}$Be $\rightarrow$ $^{10}$Be. A comparable branching ratio of $\beta p$ \cite{ayyad2019} was also observed in a recent measurement of decaying protons from $^{11}$Be stopped in the Active Target Time Projection Chamber (AT-TPC) \cite{ayyad2018}. However, various theoretical calculations \cite{borge2013,baye2011} have shown that such a $\beta p$ decay branch can only contribute to the branching ratio of $^{11}$Be $\rightarrow$ $^{10}$Be on the order of $10^{-8}$.

One possible explanation using the new $n-n'$ oscillation model is to take into account the quasi-free oscillation process in the halo, which can result in a branching ratio \cite{tan2019},
\begin{equation}
BR = \sin^2(2\theta) <\sin^2(\frac{\Delta_{nn'}}{2}\tau)> \sim 10^{-5}
\end{equation}
where $\tau$ is the $^{11}$Be lifetime. The result is in remarkable agreement with the AMS measurement. The virtual mirror neutron in the halo can undergo resonant oscillation near the edge of the nuclear potential well where $V_{eff} \sim \Delta_{nn'}$ and meanwhile overcome the barrier of the penetrability. The oscillatory $n$ and $n'$ can not escape freely due to energy conservation but will eventually decay via $\beta p$ and $\beta' p'$, respectively. The AMS approach measures the sum of $\beta p$ and $\beta' p'$ branching ratios while the AT-TPC detects only the $\beta p$ branch. In a more sensitive TPC experiment, measurement of recoiled $^{10}$Be nuclei will give a summed branching ratio of both $\beta p$ and $\beta' p'$. At the same time detection of protons will provide the partial branching ratio of $\beta p$ only. Any difference in the two ratios will indicate the existence of the $\beta' p'$ decay supporting the idea of $n-n'$ oscillations. Other possible candidates for similar decays are $^{17}$C, $^{19}$C, and $^{31}$Ne under the same criteria of one-neutron halo nuclei with $S_n < 782$ keV.

\section{Conclusion\label{conclusion}}

The above-discussed experiments can test the mirror-matter theory in different systems of neutrons, mesons, and nuclei using the current technology. The underlying model parameters, i.e., the new mixing elements of $V_{qq'}$ and mass splitting parameters can be further constrained and the consistency between them can be tested as well.
Here we summarize the proposed laboratory experiments that can test the new model \cite{tan2019} and also distinguish it from other models:
\begin{enumerate}
\item UCN decays in magnetic traps with different geometries for different mean free flight times,
\item decays of cold neutrons through strong magnetic fields (e.g. $B \sim 50$ T),
\item decays of cold neutrons in scintillation detectors made of liquid $^4$He, heavy water, or other nearly absorption-free dense materials,
\item branching fractions of $K^0_L$ and $K^0_S$ invisible decays,
\item better measurements of $K_{l3}$, $K^{\pm}_{\mu 2}$, $\pi^{\pm}_{\mu 2}$, and $\pi$ beta decays combined with better lattice QCD calculations,
\item and branching ratios of $\beta p/\beta' p'$ decays of $^{11}$Be and other one-neutron halo nuclei like $^{17}$C, $^{19}$C, and $^{31}$Ne.
\end{enumerate}

\begin{acknowledgments}
I'd like to thank Wolfgang Mittig for discussions on decay of $^{11}$Be and Yuri Kamyshkov for information on recent work. I am also grateful to Austin Reid for informing me Craig Huffer's dissertation work on lifetime measurements with liquid $^4$He and to Robert Golub for enlightening discussions on UCN in liquid helium.
This work is supported in part by the National Science Foundation under
grant No. PHY-1713857 and the Joint Institute for Nuclear Astrophysics (JINA-CEE, www.jinaweb.org), NSF-PFC under grant No. PHY-1430152.
\end{acknowledgments}

\bibliography{ckm}

\end{document}